\documentclass[aps,prl,twocolumn,superscriptaddress,floatfix]{revtex4}

\usepackage{bm}
\usepackage{amssymb}
\usepackage{graphicx}
\usepackage{amsmath}
\usepackage{dcolumn}
\usepackage[pdftex]{epsfig}
\usepackage{color}
\usepackage[german,english]{babel}
\usepackage{psfrag}
\usepackage{hyperref}
\usepackage[caption=false]{subfig}
\usepackage{longtable}
\usepackage{psfrag}
\usepackage{color}
\usepackage{float}

\definecolor{myRed}{rgb}{0.8, 0.2, 0.2}

\begin{document}

\title{Phase Diagram of Electron Doped Dichalcogenides}

\author{M. R{\"o}sner}
\email{mroesner@itp.uni-bremen.de}
\affiliation{Institut f{\"u}r Theoretische Physik, Universit{\"a}t Bremen, Otto-Hahn-Allee 1, 28359 Bremen, Germany}
\affiliation{Bremen Center for Computational Materials Science, Universit{\"a}t Bremen, Am Fallturm 1a, 28359 Bremen, Germany}

\author{S. Haas}
\affiliation{Department of Physics and Astronomy, University of Southern California, Los Angeles, CA 90089-0484, USA}

\author{T. O. Wehling}
\affiliation{Institut f{\"u}r Theoretische Physik, Universit{\"a}t Bremen, Otto-Hahn-Allee 1, 28359 Bremen, Germany}
\affiliation{Bremen Center for Computational Materials Science, Universit{\"a}t Bremen, Am Fallturm 1a, 28359 Bremen, Germany}

\date{\today}

\begin{abstract}
Using first principle calculations, we examine the sequence of phases in electron doped dichalcogenides, such as recently realized in field-gated $\rm MoS_2$. Upon increasing the electron doping level, we observe a succession of semiconducting, metallic, superconducting and charge density wave regimes, i.e. in different order compared to the phase diagram of metallic dichalcogenides such as TiSe$_2$. Both instabilities trace back to a softening of phonons which couple the electron populated conduction band minima. The superconducting dome, calculated using Eliashberg theory, is found to fit the experimentally observed phase diagram, obtained from resistivity measurements. The charge density wave phase at higher electron doping concentrations as predicted from instabilities in the phonon modes is further corroborated by detecting the accompanying lattice deformation in density functional based supercell relaxations. Upon charge density wave formation, doped MoS$_2$ remains metallic but 
undergoes a Lifschitz transition, where the number of Fermi pockets is reduced. 
\end{abstract}


\maketitle

{\it Introduction:}
Several materials including graphene or transition metal dichalcogenides can be prepared at monolayer thickness \cite{Novoselov26072005}. Because of their low effective dimensionality, there is a lack of screening in these materials, and in addition the band structure shows strong van Hove singularities. This can lead to strong enhancements of scales and result in competing instabilities, such as superconductivity (SC) and charge density wave (CDW) phases \cite{rahnejat_charge_2011,wang_electronics_2012}. 
Thereby, the quasi-two-dimensional structure of these compounds allows for a high degree of control via tuning knobs such as pressure, strain, doping and adsorbates, but it also makes these materials more vulnerable to the effects of impurity disorder. 

The generic phase diagram of the metallic transition metal dichalcogenides features a CDW regime at and close to half-filling, which upon \emph{hole doping} or exerting external pressure is suppressed by a competing SC instability \cite{PhysRevLett.86.4382,PhysRevB.83.024502}. For example, pristine $\rm 1T-TiSe_2$ undergoes a CDW phase transition at approximately 200K \cite{PhysRevB.14.4321}. Upon hole doping via Cu intercalation \cite{morosan} or application of  pressure \cite{PhysRevLett.103.236401} this phase is suppressed and replaced by competing SC order with transition temperatures  $\sim$ 2-5K, leading to a phase diagram topology akin to the high-$T_c$ cuprates, with CDW taking the place of the antiferromagnetic insulator regime in the cuprates. This succession of phases can be modeled by combining first principle calculations with Eliashberg theory, based on a phonon mediated pairing mechanism \cite{PhysRevLett.106.196406}. 

In this paper, we focus on the phase diagram of {\it electron doped} dichalcogenides. Since these matertials do not show an electron/hole symmetry it is a priori not known which phases will arise and how they compete with each other. Indeed, we find a different topology in the elctron doped regime, thus leading to an interesting set of predictions that can be experimentally tested. Without loss of generality, we focus on the much studied compound $\rm MoS_2$ because there already is a wealth of data available which allows to scrutinize our approach. 
Electron doping of thin-flake $\rm MoS_2$ has recently been achieved  by means of combined liquid/solid high-capacitance gates, leading to effective 2D carrier densities of up to $n_\text{2D} \approx 1.5\times 10^{14}\,$ cm$^{-2}$. Such doping by field effect gates allows us to access larger carrier concentrations compared to chemical substitution, without substantially deforming the lattice \cite{Ye30112012}. A field-doping-induced superconducting dome was found with onset at $n_\text{2D} = 6.8\times 10^{13}\,$ cm$^{-2}$ and peak with maximum $T_c = 10.8\,$K at $n_\text{2D} = 1.2\times 10^{14}\,$ cm$^{-2}$. \cite{Ye30112012,Hidenori_2012}. Using density functional theory calculations, it has been shown that this  superconducting dome is consistent with electron-phonon coupling that is doping-dependent due to the change of Fermi surface topology when negative charge carriers are introduced \cite{PhysRevB.87.241408}. 
Here, we push this analysis further and deliver a quantitative description of the superconducting dome and identify a competing CDW phase which occurs at higher doping concentrations. Although this kind of competition is known in the hole dope regime, it is quite surprising that the CDW phase exists in the electron doped regime as well. In this case the Fermi surface topology is totally different and thus the behavior of the newly found CDW phase is different to the corresponding phase in the hole doped case. While it may turn out to be difficult to achieve such high doping concentrations in $\rm MoS_2$ experimentally by back gating \cite{Ye30112012}, this prediction is a generic feature, and thus should hold for other electron doped dichalcogenides as well. Example systems for observing the CDW phase predicted here include chemically doped MoS$_2$, as e.g. realized by alkali deposition/intercalation \cite{Somoano_JCP_1975}.

{\it Method:} We use the VASP \cite{vasp1,vasp2} and Quantum Espresso \cite{QE-2009} Packages for the density functional theory (DFT) based self-consistent evaluation of the electronic and phononic band structures. Electron doping $x$ (in electrons per primitive MoS$_2$ unit cell) or $n_{\text{2D}} = x / A$ (in electrons per cm$^2$, $A$ is the area of the unit cell) is realized by introducing additional electrons along with a compensating jellium background. Care is taken such that no unphysical low energy states are introduced by the positive background charges. The electron-phonon matrix elements are calculated using the Phonon package of Quantum Espresso \footnote{
  The DFT calculations are performed within the LDA using norm-conserving pseudopotentials. For the electronic calculations a $32\times32\times1$ k-mesh is used ($64\times64\times1$ for the calculation of $N(\epsilon_F)$), in combination with a Methfessel-Paxton smearing ($0.0075\,$Ry). The lattice parameter is chosen to be $3.122\,$\AA\, and adjacent layers are separated by $\approx13\,$\AA. The geometry (S positions) of the simple unit cell is optimized for each electron doping. Phonon band structures and electron-phonon couplings are calculated within the density functional pertubation theory based on the evaluation of the dynamical matrices on a $8\times8\times1$ q-mesh. 
}, and the superconducting properties based on Eliashberg theory are obtained via postprocessing \cite{PhysRevB.12.905,eliashberg}. In particular, the Eliashberg spectral function,
\begin{equation}
\alpha^2 F(\omega ) = \frac{1}{2\pi N(\epsilon_F )} \sum_{{\bf q} \nu} \delta (\omega - \omega_{{\bf q}\nu} )\frac{\gamma_{{\bf q}\nu}}{\hbar \omega_{{\bf q}\nu}},
\end{equation} 
is evaluated from the electronic density of states at the Fermi level $N(\epsilon_F )$, the phonon frequencies $\omega_{{\bf q}\nu}$ and the line widths $\gamma_{{\bf q}\nu}$ which contain the electron-phonon coupling matrix elements \cite{eliashberg}. The superconducting transition temperatures can then be estimated using the Allen-Dynes formula \cite{PhysRevB.12.905},
\begin{equation} 
T_c = \frac{\hbar \omega_{log}}{1.2k_B} \exp \left[ \frac{-1.04(1+\lambda)}{\lambda (1-0.62\mu^*)-\mu^*} \right],
\end{equation}
where  $\lambda = 2\int d\omega \ \alpha^2 F(\omega)/\omega$,  $\omega_{log} = \exp [2/\lambda \int d\omega \ \alpha^2 F(\omega ) \log(\omega)/\omega ] $, and $\mu^*$ is the effective Coulomb pseudopotential. The newly found emerging CDW at higher electron concentrations is identified by (i) the occurrence of an unstable phonon mode, (ii) by spontaneous deformation of the honeycomb lattice, as well as (iii) by comparison of energies of the deformed lattice with the unperturbed lattice \footnote{ 
  The relaxed structures and the total energies of $1\times1$ and $2\times1$ supercells for several doping concentrations are calulated within the LDA. Both calculations are performed on $32\times32\times1$ k-meshes ($16\times32\times1$ in the latter case). The tetrathedron method is applied to obtain accurate total energies.
}. 

\begin{figure}[h]

  \psfrag{=}[][][0.8]{$\approx$}
  
  \psfrag{G}[][][0.9]{$\Gamma$}
  \psfrag{M}[][][0.9]{M}
  \psfrag{K}[][][0.9]{K}
  \psfrag{S}[][][0.9]{$\Sigma$}
  
  \psfrag{wph (cm-1)}[][][0.9]{\textsf{$\omega_\text{ph}$ (cm$^{-1}$)}}
  \psfrag{Eel (eV)}[][][0.9]{\textsf{$E_\text{el}$ (eV)}}
  
  \psfrag{x=0.000}[c][][0.8]{$\quad x=0.000$}
  \psfrag{x=0.025}[c][][0.8]{$\quad x=0.025$}
  \psfrag{x=0.100}[c][][0.8]{$\quad x=0.100$}
  \psfrag{x=0.150}[c][][0.8]{$\quad x=0.150$}
  
  \epsfig{file=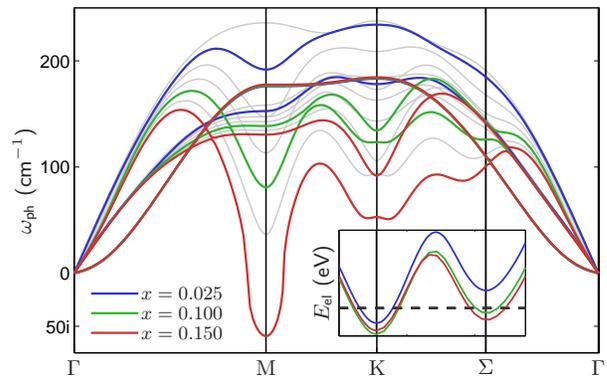,width=8.0cm}

  \caption{(Color online) Acoustic part of phononic band structure of $\rm MoS_2$ for electron doping concentrations $x=0.025$ (blue), $x=0.100$ (green), and $x=0.150$ (red), corresponding to the metallic, superconducting and charge density wave regimes respectively. The inset shows the conduction bands of the corresponding electronic band structures (Fermi levels are indicated by dashed lines, ticks are separated by $100\,$meV). The phononic band structures are completed by several doping levels inbetween the range of $x=0$ to $x=0.15$ (grey).}
  \label{bandstructure}

\end{figure}

{\it Results:}
Similar to other dichalcogenides, the low-energy properties in MoS$_2$ are dominated by minima in the conduction band at lattice vectors K and $\Sigma$, which have predominantly Mo $\rm d_{z^2}$-orbital (at K) and Mo $\rm d_{xy}$- and $\rm d_{x^2-y^2}$-orbital (at $\Sigma$) character \cite{PhysRevB.83.245213}. Upon electron doping, the $\Sigma$ valley moves towards lower energies, whereas the K valley is less affected (see inset of Fig. \ref{bandstructure}) \cite{PhysRevB.87.241408}. The doping levels shown here correspond to the metallic regime, the SC phase, and the CDW phase. These instabilities are discussed in more detail below. 

The acoustic parts of the phonon dispersions of $\rm MoS_2$ are shown for the same electron doping concentrations in Fig. \ref{bandstructure}. Similar to graphene, pristine $\rm MoS_2$ has one quadratic and two linear acoustic phonon branches which flatten out around the K and M points in an energy window $\sim 180-240\,\text{cm}^{-1}$. Upon doping, the acoustic in-plane branches soften. The parabolic out-of-plane phonons are odd under mirror transformation with respect to the Mo plane and do not couple the conduction band minima at K and $\Sigma$. There is thus no Kohn anomaly (or related phenomena) leading to softening of these phonons upon electron doping. 
At a critical electron concentration $x_c\sim0.14$ one of the acoustic modes develops an instability at the M point, indicating the onset of a CDW regime. At this point the phonon frequency of this mode becomes imaginary (Fig. \ref{bandstructure}) \cite{PhysRevLett.106.196406}.  This behavior is reminiscent of $\rm TiSe_2$, where a CDW-SC transition can be tuned by pressure or Cu intercalation \cite{PhysRevLett.103.236401}. However, the CDW regime in $\rm TiSe_2$  already occurs in its pristine state, and is suppressed by pressure or hole-doping, giving way to SC, whereas the sequence of phases we observe in  $\rm MoS_2$ is reversed. 

\begin{figure}[t]

  \psfrag{wph (cm-1)}[c][][0.9]{\textsf{$\omega_\text{ph}$ (cm$^{-1}$)}}
  \psfrag{electron doping x}[c][][0.8]{\textsf{electron doping $x$}}

  \psfrag{Phonon DOS (a.u.)}[][][0.9]{\textsf{Phonon DOS (a.u.)}}
  \psfrag{a2 F(w) (a.u.)}[][][0.9]{ \textsf{$\alpha^2 F(\omega_\text{ph})$ (a.u.)}}
  \psfrag{L(x)}[][][0.9]{$\lambda(x)$}
  \psfrag{wlog(x)}[][][0.9]{\textcolor{myRed}{$\omega_{log}(x)$}}

  \psfrag{x = 0.125}[c][][0.8]{$\quad x=0.125$}
  \psfrag{x = 0.112}[c][][0.8]{$\quad x=0.112$}
  \psfrag{x = 0.100}[c][][0.8]{$\quad x=0.100$}
  \psfrag{x = 0.087}[c][][0.8]{$\quad x=0.087$}
  \psfrag{x = 0.075}[c][][0.8]{$\quad x=0.075$}
  
  \epsfig{file=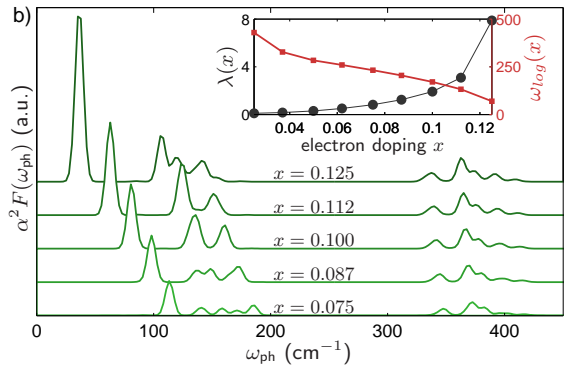,width=7.5cm}
  
  \caption{(Color online) Eliashberg function of $\rm MoS_2$ in the SC phase for different doping levels. The inset shows the evolution $\lambda(x)$ and $\omega_{log}(x)$ (using a Gaussian smearing of $\delta = 0.005\,$Ry), determining $T_c(x)$.}
  \label{dynamics}
  
\end{figure}

Let us now turn our focus toward the SC regime at intermediate concentrations. We  examine the lattice dynamics encoded in the phonon density of states and the Eliashberg function. The phonon density of states (see Suppl. Mat.) has a rich peak structure, with the largest contributions stemming from the regions where the phonon dispersion flattens, leading to characteristic van Hove enhancements. 
The SC response encoded in the Eliashberg function is dominated by  the flat regions (around the M and K points) of the acoustic phonon branches. These features are inherited by the Eliashberg function (Fig. \ref{dynamics}), which includes weighting by the electron-phonon coupling matrix elements. 

\begin{figure}[t!]

  \psfrag{Tc (K)}[][][0.9]{\textsf{critical temperature, $T_c(x)$ (K)}}
  \psfrag{a}[][][0.9]{\textcolor{myRed}{\textsf{lattice distortion, $\alpha(x)$} ($^\circ$)}}

  \psfrag{electron doping n2D (1014 cm-2)}[c][][0.9]{\textsf{electron doping $n_{\text{2D}}$ ($10^{14}$ cm$^{-2}$)}}
  \psfrag{electron doping x}[c][][0.9]{\textsf{electron doping $x$}}
  
  \psfrag{semi-metal}[][][0.8]{\textsf{metal}}
  \psfrag{superconductor}[][][0.8]{\textsf{superconductor}}
  \psfrag{CDW}[][][0.8]{\textsf{CDW}}
  
  \psfrag{deformation}[l][l][0.7]{\textsf{$\alpha(x)$}}
  \psfrag{mu=0.05}[l][l][0.7]{\textsf{$\mu^* = 0.05$}}
  \psfrag{mu=0.15}[l][l][0.7]{\textsf{$\mu^* = 0.15$}}
  \psfrag{mu=0.25}[l][l][0.7]{\textsf{$\mu^* = 0.25$}}
  \psfrag{CDW formation}[l][l][0.7]{\textsf{CDW formation}}
  \psfrag{energy}[l][l][0.7]{\textsf{energy}}
  \psfrag{exp. data}[l][l][0.7]{\textsf{exp. data}}
  
  \psfrag{a05877}[l][l][0.7]{\textsf{$a=3.110$\AA}}
  \psfrag{a05900}[l][l][0.7]{\textsf{$a=3.122$\AA}}
  \psfrag{a05922}[l][l][0.7]{\textsf{$a=3.134$\AA}}
  \psfrag{a05944}[l][l][0.7]{\textsf{$a=3.146$\AA}}
  
  \epsfig{file=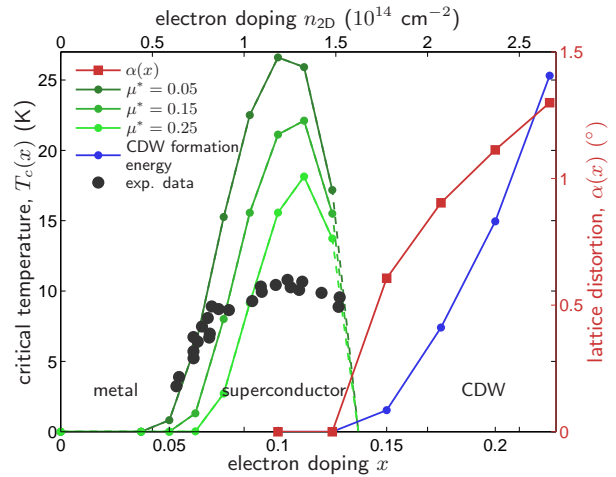,width=8.0cm}
  
  \caption{(Color online) Temperature-doping phase diagram of $\rm MoS_2$. Circles belong to the left axis (K) and squares to the right ($^\circ$). Green lines are obtained from first principle calculations combined with Eliashberg theory and show the SC critical temperature for different Coulomb pseudopotentials(using a Gaussian smearing of $\delta = 0.005\,$Ry). Dashed lines are guides to the eye. Black circles are experimental data \cite{Ye30112012}. Also shown is the lattice distortion angle $\alpha$ (red squares) and the energy gain (blue circles) upon CDW formation.}
  \label{dopingdependence}

\end{figure}

As the acoustic phonon mode with minimum at the M point softens, the evolution of the Eliashberg function displays a maximum integrated intensity at $x=0.125$. However, this concentration does not correspond to maximum of $T_c(x)$ since the interplay of the effective coupling $\lambda(x)$ and $\omega_{log}(x)$ has to be considered. As it can be seen in the inset of  Fig. \ref{dynamics} $\omega_{log}(x)$ decreases while $\lambda(x)$ increases with increasing doping. An optimal proportion is reached at $x\approx 0.11$ leading to a maximum of $T_c$. 
Thus, the combined evolution of $\lambda(x)$ and $\omega_{log}(x)$ is one reason for the dome-shaped dependence of the SC transition temperature on the electron doping concentration, which can be seen in Fig. \ref{dopingdependence}. Here we show experimental data of Ref. \cite{Ye30112012} along with results of our numerical simulation for different Coulomb pseudopotentials $\mu^*$. 
Besides the coincidence in the position of the maximum in $T_c(x)$ at $x\approx0.11$ ($n_{\text{2D}} = 1.2\times 10^{14}\,$ cm$^{-2}$), we also note that the computed and experimental SC transition temperatures are of the same order of magnitude. This is remarkable, since Eliashberg theory is a rather crude approximation, which does not account for pair-breaking effects, such as impurities, incorporates Coulomb interactions only statically as $\mu^*$ and neglects enhanced phase fluctuations in 2D. It is therefore expected to overestimate $T_c(x)$. But, this overestimation can be reduced by involving a doping-dependent $\mu^*(x)$, which is reasonable, since the Coulomb interaction will clearly change upon electron doping. And indeed, we find an even better quantitative agreement with the experimental data, by adjusting $\mu^*$ in dependence of the doping concentration, as it can be seen Fig. \ref{dopingdependence}. Hence, we find a second reason for the dome-shape of the $T_c(x)$.

\begin{figure*}[ht]
  
  \epsfig{file=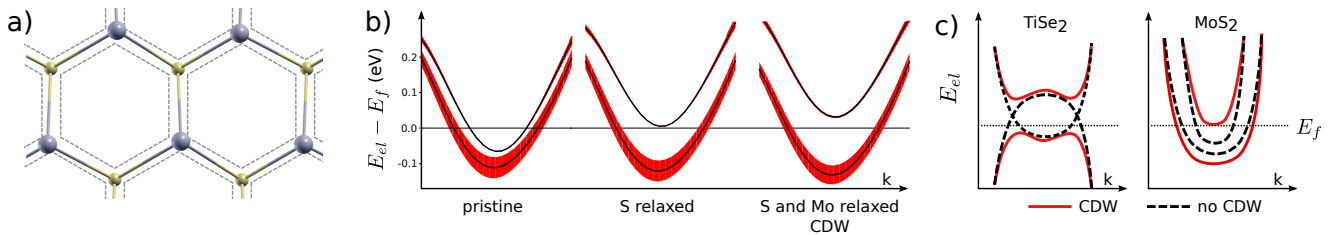,height=3.0cm}
  
  \caption{(Color online) Manifestation of the CDW in lattice distortion and band structure. (a) Lattice distortion in $\rm MoS_2$, observed in ab initio calculations via self-consistent relaxation of 2$\times$1 supercells. (b) Influence of lattice relaxation effects on the band structures of the 2$\times$1 supercells obtained for a doping level $x=0.2$. The $\rm d_{xy},d_{x^2-y^2}$ weight of the bands is illustrated by the (red) width of the bands. The middle panel shows the effect of homogeneous outward relaxation of the S atoms on the conduction band minima as observable from comparison of the $\rm MoS_2$ structure with S positions according to undoped (pristine) and homogeneously relaxed $x=0.2$ system. The right panel shows the comparison of supercell band structures for homogeneously relaxed $\rm MoS_2$ and $\rm MoS_2$ featuring the CDW (fully relaxed) at $x=0.2$. 
  (c) Comparison of band topologies involved in the CDW formation in $\rm TiSe_2$ and $\rm MoS_2$. Only in the $\rm TiSe_2$ case, CDW formation can fully gap the Fermi surface.}
  \label{distortion}
  
\end{figure*}

In order to better understand the nature of the  SC-CDW phase transition, we examine the doping dependence of the CDW-induced lattice distortion $\alpha$, shown as a red line in Fig. \ref{dopingdependence}. Here,  $\alpha$ is defined as the angle between three neighboring Mo atoms subtracted by $60^\circ$. For an undistorted honeycomb lattice one finds $\alpha=0$. By relaxing the atomic structure of a 2$\times$1 supercell, we observe $\alpha \ne 0$ beyond a critical electron concentration of $x_c \approx 0.14$, as forces arise from the unstable phonon mode at lattice vector $M$ \cite{ge2014}. These distortion effects, depicted in Fig. \ref{distortion} (a),  become more pronounced with increasing electron doping. We note that in addition to the CDW formation, there is a further homogeneous outward relaxation of the S atoms upon electron doping.

The effects of homogeneous S relaxation and CDW formation on the electronic structure are illustrated in Fig. \ref{distortion} (b) for electron doping  $x=0.2$  \footnote{For these calculations the PAW method in the LDA as implemented in VASP has been used.}. In the supercell Brillouin zone, the former band minima at K and $\Sigma$ are folded almost on top of each other at the supercell K point. In the absence of a CDW, low-energy states originating from K and $\Sigma$ can be distinguished by their orbital band character. The latter states carry a significant $\rm d_{xy},d_{x^2-y^2}$-weight, whereas the conduction band minimum at K has no such admixture (see Fig. \ref{distortion} (b) left panel). The outward relaxation of the S atoms lowers $\rm d_{xy},d_{x^2-y^2}$-derived states from $\Sigma$ in energy (Fig. \ref{distortion} (b) middle panel). 
With increasing CDW amplitude the two bands originating from K and $\Sigma$ mix, and this hybridization adds to the splitting of the two bands, c.f. Fig. \ref{distortion} (b) (right panel). This splitting leads to lowering of the electronic energy if the Fermi level lies sufficiently high in the conduction band. The total energy gain upon CDW formation as function of doping level is shown in Fig. \ref{dopingdependence} (blue line). It illustrates that the CDW formation energies for $x<0.25$ are comparable to typical Cooper pair condensation energies $\sim 10$\,K ($\sim 1$\,meV) encountered here, and an interesting competition of the two should emerge.

While the Eliashberg theory of SC order is only applicable as long as the lattice remains stable, it is clear that the competition of CDW and SC order will in any case  depend on changes of the Fermi surface due to CDW formation. For the perfect crystal (relaxed structure at zero doping) and a doping level of $x=0.2$, two bands would intersect the Fermi level near the supercell K point, and there would be thus two Fermi lines around K. Upon outward relaxation of the S atoms (preserving all lattice symmetries) and formation of the CDW, we observe a Lifschitz transition where one of the Fermi pockets disappears, Fig. \ref{distortion} (b) (middle and right panel). The system thus remains metallic in the CDW phase, but the SC transition temperatures should be reduced due to the vanishing phase space for inter-pocket scattering. Persisting metallicity in the CDW phase of MoS$_2$ is indeed ensured by the ``topology'' of the inter-mixing bands at K and $\Sigma$, Fig. \ref{distortion} (c). 
In  TiSe$_2$, CDW bands with opposite slope are folded on top of each other, and a gap can open upon hybridization. However, in MoS$_2$, the slopes of the backfolded bands have the same sign, and avoided crossings do not lead to a full gap, but only reduce the number of Fermi sheets by one.

Since it is known, that the energies of the minima in the counduction band of MoS$_2$ are very sensitive to external strain \cite{acs_nanolett} the before mentioned change in the Fermi surface topology due to the CDW transition might be sensitive to strain as well. To analyse this behavior we redid the relaxation calculations in the doping range of the inset of the charge density wave for different lattice constants (see Suppl. Mat.). Thereby we found a strong dependence of the critical doping concentration at which the CDW phase sets in to the lattice constant. A strain of less the $1\%$ changes the critical doping concentration by more than $15\%$. Thus, the competition between the CDW and the SC phase can be triggered and tuned drastically by both, the doping concentration and the external strain. 

{\it Conclusions:} Electron doped dichalcogenides feature CDW and SC instabilities, driven by the softening of an acoustical phonon mode upon charge doping.
Due to the band topology, the M point CDW cannot fully gap the Fermi surface of electron doped MoS$_2$. Therefore, CDW and SC phases may coexist, albeit with reduced SC transition temperatures. 
In any case, the SC and CDW instabilities rely on the energy differences between the conduction band minima at K and $\Sigma$. These are highly sensitive to lattice relaxation, and we speculate that adsorption of molecular species on MoS$_2$ may be useful for tuning superconducting transition temperatures. 

The competition of CDW and SC phases is common in metallic transition metal dichalcogenides, such as TiSe$_2$, NbSe$_2$ and  TaS$_2$. All these materials differ, however, from MoS$_2$ in that the transition metal atoms lack one (Nb, Ta) or two (Ti) valence electrons in comparison to Mo. Nevertheless, electron doped MoS$_2$ develops CDW/SC instabilities as well, although entirely different bands are involved. The most prominent resulting difference compared to materials like TiSe$_2$ is the reversed order in the phase diagram of MoS$_2$.

{\it Acknowledgments:} We are grateful for useful discussions with A.V. Balatsky, I. Gierz, A. Liu, F. Mauri, and M. Calandra. S.H. would like to the thank the Humboldt Foundation for support. This work was supported by the European Graphene Flagship and by the Department of Energy under Grant No. DE-FG02-05ER46240. The numerical computations were carried out on the University of Southern California high performance supercomputer cluster and the Norddeutscher Verbund zur F\"{o}rderung des Hoch- und H\"{o}chstleistungsrechnens (HLRN) cluster. 

\bibliography{references}

\providecommand{\noopsort}[1]{}\providecommand{\singleletter}[1]{#1}%
\begin{thebibliography}{10}%
\makeatletter
\providecommand \@ifxundefined [1]{%
 \ifx #1\undefined \expandafter \@firstoftwo
 \else \expandafter \@secondoftwo
\fi
}%
\providecommand \@ifnum [1]{%
 \ifnum #1\expandafter \@firstoftwo
 \else \expandafter \@secondoftwo
\fi
}%
\providecommand \enquote [1]{``#1''}%
\providecommand \bibnamefont  [1]{#1}%
\providecommand \bibfnamefont [1]{#1}%
\providecommand \citenamefont [1]{#1}%
\providecommand\href[0]{\@sanitize\@href}%
\providecommand\@href[1]{\endgroup\@@startlink{#1}\endgroup\@@href}%
\providecommand\@@href[1]{#1\@@endlink}%
\providecommand \@sanitize [0]{\begingroup\catcode`\&12\catcode`\#12\relax}%
\@ifxundefined \pdfoutput {\@firstoftwo}{%
 \@ifnum{\z@=\pdfoutput}{\@firstoftwo}{\@secondoftwo}%
}{%
 \providecommand\@@startlink[1]{\leavevmode\special{html:<a href="#1">}}%
 \providecommand\@@endlink[0]{\special{html:</a>}}%
}{%
 \providecommand\@@startlink[1]{%
  \leavevmode
  \pdfstartlink
   attr{/Border[0 0 1 ]/H/I/C[0 1 1]}%
   user{/Subtype/Link/A<</Type/Action/S/URI/URI(#1)>>}%
  \relax
 }%
 \providecommand\@@endlink[0]{\pdfendlink}%
}%
\providecommand \url  [0]{\begingroup\@sanitize \@url }%
\providecommand \@url [1]{\endgroup\@href {#1}{\urlprefix}}%
\providecommand \urlprefix [0]{URL }%
\providecommand \Eprint[0]{\href }%
\@ifxundefined \urlstyle {%
  \providecommand \doi [1]{doi:\discretionary{}{}{}#1}%
}{%
  \providecommand \doi [0]{doi:\discretionary{}{}{}\begingroup
  \urlstyle{rm}\Url }%
}%
\providecommand \doibase [0]{http://dx.doi.org/}%
\providecommand \Doi[1]{\href{\doibase#1}}%
\providecommand \bibAnnote [3]{%
  \BibitemShut{#1}%
  \begin{quotation}\noindent
    \textsc{Key:}\ #2\\\textsc{Annotation:}\ #3%
  \end{quotation}%
}%
\providecommand \bibAnnoteFile [2]{%
  \IfFileExists{#2}{\bibAnnote {#1} {#2} {\input{#2}}}{}%
}%
\providecommand \typeout [0]{\immediate \write \m@ne }%
\providecommand \selectlanguage [0]{\@gobble}%
\providecommand \bibinfo [0]{\@secondoftwo}%
\providecommand \bibfield [0]{\@secondoftwo}%
\providecommand \translation [1]{[#1]}%
\providecommand \BibitemOpen[0]{}%
\providecommand \bibitemStop [0]{}%
\providecommand \bibitemNoStop [0]{.\EOS\space}%
\providecommand \EOS [0]{\spacefactor3000\relax}%
\providecommand \BibitemShut [1]{\csname bibitem#1\endcsname}%
\bibitem{Novoselov26072005}%
  \BibitemOpen
  \bibfield{author}{%
  \bibinfo {author} {\bibfnamefont{K.~S.}\ \bibnamefont{Novoselov}}, \bibinfo
  {author} {\bibfnamefont{D.}~\bibnamefont{Jiang}}, \bibinfo {author}
  {\bibfnamefont{F.}~\bibnamefont{Schedin}}, \bibinfo {author}
  {\bibfnamefont{T.~J.}\ \bibnamefont{Booth}}, \bibinfo {author}
  {\bibfnamefont{V.~V.}\ \bibnamefont{Khotkevich}}, \bibinfo {author}
  {\bibfnamefont{S.~V.}\ \bibnamefont{Morozov}},\ and\ \bibinfo {author}
  {\bibfnamefont{A.~K.}\ \bibnamefont{Geim}},\ }%
  \bibfield{journal}{%
  \Doi{10.1073/pnas.0502848102}{\bibinfo {journal} {Proc. Natl. Acad. Sci.
  USA}}\ }%
  \textbf{\bibinfo {volume} {102}},\ \bibinfo {pages} {10451} (\bibinfo {year}
  {2005})%
  \bibAnnoteFile{NoStop}{Novoselov26072005}%
\bibitem{rahnejat_charge_2011}%
  \BibitemOpen
  \bibfield{author}{%
  \bibinfo {author} {\bibfnamefont{K.~C.}\ \bibnamefont{Rahnejat}}, \bibinfo
  {author} {\bibfnamefont{C.~A.}\ \bibnamefont{Howard}}, \bibinfo {author}
  {\bibfnamefont{N.~E.}\ \bibnamefont{Shuttleworth}}, \bibinfo {author}
  {\bibfnamefont{S.~R.}\ \bibnamefont{Schofield}}, \bibinfo {author}
  {\bibfnamefont{K.}~\bibnamefont{Iwaya}}, \bibinfo {author}
  {\bibfnamefont{C.~F.}\ \bibnamefont{Hirjibehedin}}, \bibinfo {author}
  {\bibfnamefont{C.}~\bibnamefont{Renner}}, \bibinfo {author}
  {\bibfnamefont{G.}~\bibnamefont{Aeppli}},\ and\ \bibinfo {author}
  {\bibfnamefont{M.}~\bibnamefont{Ellerby}},\ }%
  \bibfield{journal}{%
  \Doi{10.1038/ncomms1574}{\bibinfo {journal} {Nature Communications}}\ }%
  \textbf{\bibinfo {volume} {2}},\ \bibinfo {pages} {558} (\bibinfo {month}
  {nov}\ \bibinfo {year} {2011})%
  \bibAnnoteFile{NoStop}{rahnejat_charge_2011}%
\bibitem{wang_electronics_2012}%
  \BibitemOpen
  \bibfield{author}{%
  \bibinfo {author} {\bibfnamefont{Q.~H.}\ \bibnamefont{Wang}}, \bibinfo
  {author} {\bibfnamefont{K.}~\bibnamefont{Kalantar-Zadeh}}, \bibinfo {author}
  {\bibfnamefont{A.}~\bibnamefont{Kis}}, \bibinfo {author}
  {\bibfnamefont{J.~N.}\ \bibnamefont{Coleman}},\ and\ \bibinfo {author}
  {\bibfnamefont{M.~S.}\ \bibnamefont{Strano}},\ }%
  \bibfield{journal}{%
  \Doi{10.1038/nnano.2012.193}{\bibinfo {journal} {Nature Nanotechnology}}\ }%
  \textbf{\bibinfo {volume} {7}},\ \bibinfo {pages} {699} (\bibinfo {year}
  {2012}),\ ISSN \bibinfo {issn} {1748-3387}%
  \bibAnnoteFile{NoStop}{wang_electronics_2012}%
\bibitem{PhysRevLett.86.4382}%
  \BibitemOpen
  \bibfield{author}{%
  \bibinfo {author} {\bibfnamefont{A.~H.}\ \bibnamefont{Castro~Neto}},\ }%
  \bibfield{journal}{%
  \Doi{10.1103/PhysRevLett.86.4382}{\bibinfo {journal} {Phys. Rev. Lett.}}\ }%
  \textbf{\bibinfo {volume} {86}},\ \bibinfo {pages} {4382} (\bibinfo {month}
  {May}\ \bibinfo {year} {2001})%
  \bibAnnoteFile{NoStop}{PhysRevLett.86.4382}%
\bibitem{PhysRevB.83.024502}%
  \BibitemOpen
  \bibfield{author}{%
  \bibinfo {author} {\bibfnamefont{J.}~\bibnamefont{van Wezel}}, \bibinfo
  {author} {\bibfnamefont{P.}~\bibnamefont{Nahai-Williamson}},\ and\ \bibinfo
  {author} {\bibfnamefont{S.~S.}\ \bibnamefont{Saxena}},\ }%
  \bibfield{journal}{%
  \Doi{10.1103/PhysRevB.83.024502}{\bibinfo {journal} {Phys. Rev. B}}\ }%
  \textbf{\bibinfo {volume} {83}},\ \bibinfo {pages} {024502} (\bibinfo {month}
  {Jan}\ \bibinfo {year} {2011})%
  \bibAnnoteFile{NoStop}{PhysRevB.83.024502}%
\bibitem{PhysRevB.14.4321}%
  \BibitemOpen
  \bibfield{author}{%
  \bibinfo {author} {\bibfnamefont{F.~J.}\ \bibnamefont{Di~Salvo}}, \bibinfo
  {author} {\bibfnamefont{D.~E.}\ \bibnamefont{Moncton}},\ and\ \bibinfo
  {author} {\bibfnamefont{J.~V.}\ \bibnamefont{Waszczak}},\ }%
  \bibfield{journal}{%
  \Doi{10.1103/PhysRevB.14.4321}{\bibinfo {journal} {Phys. Rev. B}}\ }%
  \textbf{\bibinfo {volume} {14}},\ \bibinfo {pages} {4321} (\bibinfo {month}
  {Nov}\ \bibinfo {year} {1976})%
  \bibAnnoteFile{NoStop}{PhysRevB.14.4321}%
\bibitem{morosan}%
  \BibitemOpen
  \bibfield{author}{%
  \bibinfo {author} {\bibfnamefont{E.}~\bibnamefont{Morosan}}, \bibinfo
  {author} {\bibfnamefont{H.}~\bibnamefont{Zandbergen}}, \bibinfo {author}
  {\bibfnamefont{B.}~\bibnamefont{Dennis}}, \bibinfo {author}
  {\bibfnamefont{J.}~\bibnamefont{Bos}}, \bibinfo {author}
  {\bibfnamefont{Y.}~\bibnamefont{Onose}}, \bibinfo {author}
  {\bibfnamefont{T.}~\bibnamefont{Klimczuk}}, \bibinfo {author}
  {\bibfnamefont{A.}~\bibnamefont{Ramirez}}, \bibinfo {author}
  {\bibfnamefont{N.}~\bibnamefont{Ong}},\ and\ \bibinfo {author}
  {\bibfnamefont{R.}~\bibnamefont{Cava}},\ }%
  \bibfield{journal}{%
  \bibinfo {journal} {Nat Phys}\ }%
  \textbf{\bibinfo {volume} {2}},\ \bibinfo {pages} {544} (\bibinfo {year}
  {2006})%
  \bibAnnoteFile{NoStop}{morosan}%
\bibitem{PhysRevLett.103.236401}%
  \BibitemOpen
  \bibfield{author}{%
  \bibinfo {author} {\bibfnamefont{A.~F.}\ \bibnamefont{Kusmartseva}}, \bibinfo
  {author} {\bibfnamefont{B.}~\bibnamefont{Sipos}}, \bibinfo {author}
  {\bibfnamefont{H.}~\bibnamefont{Berger}}, \bibinfo {author}
  {\bibfnamefont{L.}~\bibnamefont{Forr\'o}},\ and\ \bibinfo {author}
  {\bibfnamefont{E.}~\bibnamefont{Tuti\ifmmode~\check{s}\else \v{s}\fi{}}},\ }%
  \bibfield{journal}{%
  \Doi{10.1103/PhysRevLett.103.236401}{\bibinfo {journal} {Phys. Rev. Lett.}}\
  }%
  \textbf{\bibinfo {volume} {103}},\ \bibinfo {pages} {236401} (\bibinfo
  {month} {Nov}\ \bibinfo {year} {2009})%
  \bibAnnoteFile{NoStop}{PhysRevLett.103.236401}%
\bibitem{PhysRevLett.106.196406}%
  \BibitemOpen
  \bibfield{author}{%
  \bibinfo {author} {\bibfnamefont{M.}~\bibnamefont{Calandra}}\ and\ \bibinfo
  {author} {\bibfnamefont{F.}~\bibnamefont{Mauri}},\ }%
  \bibfield{journal}{%
  \Doi{10.1103/PhysRevLett.106.196406}{\bibinfo {journal} {Phys. Rev. Lett.}}\
  }%
  \textbf{\bibinfo {volume} {106}},\ \bibinfo {pages} {196406} (\bibinfo
  {month} {May}\ \bibinfo {year} {2011})%
  \bibAnnoteFile{NoStop}{PhysRevLett.106.196406}%
\bibitem{Ye30112012}%
  \BibitemOpen
  \bibfield{author}{%
  \bibinfo {author} {\bibfnamefont{J.~T.}\ \bibnamefont{Ye}}, \bibinfo {author}
  {\bibfnamefont{Y.~J.}\ \bibnamefont{Zhang}}, \bibinfo {author}
  {\bibfnamefont{R.}~\bibnamefont{Akashi}}, \bibinfo {author}
  {\bibfnamefont{M.~S.}\ \bibnamefont{Bahramy}}, \bibinfo {author}
  {\bibfnamefont{R.}~\bibnamefont{Arita}},\ and\ \bibinfo {author}
  {\bibfnamefont{Y.}~\bibnamefont{Iwasa}},\ }%
  \bibfield{journal}{%
  \Doi{10.1126/science.1228006}{\bibinfo {journal} {Science}}\ }%
  \textbf{\bibinfo {volume} {338}},\ \bibinfo {pages} {1193} (\bibinfo {year}
  {2012})%
  \bibAnnoteFile{NoStop}{Ye30112012}%
\bibitem{Hidenori_2012}%
  \BibitemOpen
  \bibfield{author}{%
  \bibinfo {author} {\bibfnamefont{K.}~\bibnamefont{Taniguchi}}, \bibinfo
  {author} {\bibfnamefont{A.}~\bibnamefont{Matsumoto}}, \bibinfo {author}
  {\bibfnamefont{H.}~\bibnamefont{Shimotani}},\ and\ \bibinfo {author}
  {\bibfnamefont{H.}~\bibnamefont{Takagi}},\ }%
  \bibfield{journal}{%
  \Doi{http://dx.doi.org/10.1063/1.4740268}{\bibinfo {journal} {Applied Physics
  Letters}}\ }%
  \textbf{\bibinfo {volume} {101}},\ \bibinfo {eid} {042603} (\bibinfo {year}
  {2012})%
  \bibAnnoteFile{NoStop}{Hidenori_2012}%
\bibitem{PhysRevB.87.241408}%
  \BibitemOpen
  \bibfield{author}{%
  \bibinfo {author} {\bibfnamefont{Y.}~\bibnamefont{Ge}}\ and\ \bibinfo
  {author} {\bibfnamefont{A.~Y.}\ \bibnamefont{Liu}},\ }%
  \bibfield{journal}{%
  \Doi{10.1103/PhysRevB.87.241408}{\bibinfo {journal} {Phys. Rev. B}}\ }%
  \textbf{\bibinfo {volume} {87}},\ \bibinfo {pages} {241408} (\bibinfo {month}
  {Jun}\ \bibinfo {year} {2013})%
  \bibAnnoteFile{NoStop}{PhysRevB.87.241408}%
\bibitem{Somoano_JCP_1975}%
  \BibitemOpen
  \bibfield{author}{%
  \bibinfo {author} {\bibfnamefont{R.~B.}\ \bibnamefont{Somoano}}, \bibinfo
  {author} {\bibfnamefont{V.}~\bibnamefont{Hadek}}, \bibinfo {author}
  {\bibfnamefont{A.}~\bibnamefont{Rembaum}}, \bibinfo {author}
  {\bibfnamefont{S.}~\bibnamefont{Samson}},\ and\ \bibinfo {author}
  {\bibfnamefont{J.~A.}\ \bibnamefont{Woollam}},\ }%
  \bibfield{journal}{%
  \bibinfo {journal} {The Journal of Chemical Physics}\ }%
  \textbf{\bibinfo {volume} {62}},\ \bibinfo {pages} {1068} (\bibinfo {year}
  {1975})%
  \bibAnnoteFile{NoStop}{Somoano_JCP_1975}%
\bibitem{vasp1}%
  \BibitemOpen
  \bibfield{author}{%
  \bibinfo {author} {\bibfnamefont{G.}~\bibnamefont{Kresse}}\ and\ \bibinfo
  {author} {\bibfnamefont{J.}~\bibnamefont{Furthm{\"u}ller}},\ }%
  \bibfield{journal}{%
  \bibinfo {journal} {Comput. Mat. Sci.},\ \bibinfo {pages} {15}}%
   (\bibinfo {year} {1996})%
  \bibAnnoteFile{NoStop}{vasp1}%
\bibitem{vasp2}%
  \BibitemOpen
  \bibfield{author}{%
  \bibinfo {author} {\bibfnamefont{G.}~\bibnamefont{Kresse}}\ and\ \bibinfo
  {author} {\bibfnamefont{J.}~\bibnamefont{Furthm{\"u}ller}},\ }%
  \bibfield{journal}{%
  \bibinfo {journal} {Phys. Rev. B},\ \bibinfo {pages} {11169}}%
   (\bibinfo {year} {1996})%
  \bibAnnoteFile{NoStop}{vasp2}%
\bibitem{QE-2009}%
  \BibitemOpen
  \bibfield{author}{%
  \bibinfo {author} {\bibfnamefont{P.}~\bibnamefont{Giannozzi}}, \bibinfo
  {author} {\bibfnamefont{S.}~\bibnamefont{Baroni}}, \bibinfo {author}
  {\bibfnamefont{N.}~\bibnamefont{Bonini}}, \bibinfo {author}
  {\bibfnamefont{M.}~\bibnamefont{Calandra}}, \bibinfo {author}
  {\bibfnamefont{R.}~\bibnamefont{Car}}, \bibinfo {author}
  {\bibfnamefont{C.}~\bibnamefont{Cavazzoni}}, \bibinfo {author}
  {\bibfnamefont{D.}~\bibnamefont{Ceresoli}}, \bibinfo {author}
  {\bibfnamefont{G.~L.}\ \bibnamefont{Chiarotti}}, \bibinfo {author}
  {\bibfnamefont{M.}~\bibnamefont{Cococcioni}}, \bibinfo {author}
  {\bibfnamefont{I.}~\bibnamefont{Dabo}}, \bibinfo {author}
  {\bibfnamefont{A.}~\bibnamefont{{Dal Corso}}}, \bibinfo {author}
  {\bibfnamefont{S.}~\bibnamefont{de~Gironcoli}}, \bibinfo {author}
  {\bibfnamefont{S.}~\bibnamefont{Fabris}}, \bibinfo {author}
  {\bibfnamefont{G.}~\bibnamefont{Fratesi}}, \bibinfo {author}
  {\bibfnamefont{R.}~\bibnamefont{Gebauer}}, \bibinfo {author}
  {\bibfnamefont{U.}~\bibnamefont{Gerstmann}}, \bibinfo {author}
  {\bibfnamefont{C.}~\bibnamefont{Gougoussis}}, \bibinfo {author}
  {\bibfnamefont{A.}~\bibnamefont{Kokalj}}, \bibinfo {author}
  {\bibfnamefont{M.}~\bibnamefont{Lazzeri}}, \bibinfo {author}
  {\bibfnamefont{L.}~\bibnamefont{Martin-Samos}}, \bibinfo {author}
  {\bibfnamefont{N.}~\bibnamefont{Marzari}}, \bibinfo {author}
  {\bibfnamefont{F.}~\bibnamefont{Mauri}}, \bibinfo {author}
  {\bibfnamefont{R.}~\bibnamefont{Mazzarello}}, \bibinfo {author}
  {\bibfnamefont{S.}~\bibnamefont{Paolini}}, \bibinfo {author}
  {\bibfnamefont{A.}~\bibnamefont{Pasquarello}}, \bibinfo {author}
  {\bibfnamefont{L.}~\bibnamefont{Paulatto}}, \bibinfo {author}
  {\bibfnamefont{C.}~\bibnamefont{Sbraccia}}, \bibinfo {author}
  {\bibfnamefont{S.}~\bibnamefont{Scandolo}}, \bibinfo {author}
  {\bibfnamefont{G.}~\bibnamefont{Sclauzero}}, \bibinfo {author}
  {\bibfnamefont{A.~P.}\ \bibnamefont{Seitsonen}}, \bibinfo {author}
  {\bibfnamefont{A.}~\bibnamefont{Smogunov}}, \bibinfo {author}
  {\bibfnamefont{P.}~\bibnamefont{Umari}},\ and\ \bibinfo {author}
  {\bibfnamefont{R.~M.}\ \bibnamefont{Wentzcovitch}},\ }%
  \bibfield{journal}{%
  \bibinfo {journal} {Journal of Physics: Condensed Matter}\ }%
  \textbf{\bibinfo {volume} {21}},\ \bibinfo {pages} {395502} (\bibinfo {year}
  {2009})%
  \bibAnnoteFile{NoStop}{QE-2009}%
\bibitem{PhysRevB.12.905}%
  \BibitemOpen
  \bibfield{author}{%
  \bibinfo {author} {\bibfnamefont{P.~B.}\ \bibnamefont{Allen}}\ and\ \bibinfo
  {author} {\bibfnamefont{R.~C.}\ \bibnamefont{Dynes}},\ }%
  \bibfield{journal}{%
  \Doi{10.1103/PhysRevB.12.905}{\bibinfo {journal} {Phys. Rev. B}}\ }%
  \textbf{\bibinfo {volume} {12}},\ \bibinfo {pages} {905} (\bibinfo {month}
  {Aug}\ \bibinfo {year} {1975})%
  \bibAnnoteFile{NoStop}{PhysRevB.12.905}%
\bibitem{eliashberg}%
  \BibitemOpen
  \bibfield{author}{%
  \bibinfo {author} {\bibfnamefont{G.~M.}\ \bibnamefont{Eliashberg}},\ }%
  \bibfield{journal}{%
  \bibinfo {journal} {Zh. Eksp. Teor. Fiz.},\ \bibinfo {pages} {966}}%
   (\bibinfo {year} {1960})%
  \bibAnnoteFile{NoStop}{eliashberg}%
\bibitem{PhysRevB.83.245213}%
  \BibitemOpen
  \bibfield{author}{%
  \bibinfo {author} {\bibfnamefont{A.}~\bibnamefont{Kuc}}, \bibinfo {author}
  {\bibfnamefont{N.}~\bibnamefont{Zibouche}},\ and\ \bibinfo {author}
  {\bibfnamefont{T.}~\bibnamefont{Heine}},\ }%
  \bibfield{journal}{%
  \Doi{10.1103/PhysRevB.83.245213}{\bibinfo {journal} {Phys. Rev. B}}\ }%
  \textbf{\bibinfo {volume} {83}},\ \bibinfo {pages} {245213} (\bibinfo {month}
  {Jun}\ \bibinfo {year} {2011})%
  \bibAnnoteFile{NoStop}{PhysRevB.83.245213}%
\bibitem{ge2014}%
  \BibitemOpen
  \bibfield{author}{%
  \bibinfo {author} {\bibfnamefont{Y.}~\bibnamefont{Ge}}, \bibinfo {author}
  {\bibfnamefont{W.}~\bibnamefont{Wan}}, \bibinfo {author}
  {\bibfnamefont{W.}~\bibnamefont{Feng}}, \bibinfo {author}
  {\bibfnamefont{D.}~\bibnamefont{Xiao}},\ and\ \bibinfo {author}
  {\bibfnamefont{Y.}~\bibnamefont{Yao}}\ }%
  \Eprint{http://arxiv.org/abs/arXiv:1403.0695}{arXiv:1403.0695}%
  \bibAnnoteFile{NoStop}{ge2014}%
\bibitem{acs_nanolett}%
  \BibitemOpen
  \bibfield{author}{%
  \bibinfo {author} {\bibfnamefont{A.}~\bibnamefont{Steinhoff}}, \bibinfo
  {author} {\bibfnamefont{M.}~\bibnamefont{Rösner}}, \bibinfo {author}
  {\bibfnamefont{F.}~\bibnamefont{Jahnke}}, \bibinfo {author}
  {\bibfnamefont{T.~O.}\ \bibnamefont{Wehling}},\ and\ \bibinfo {author}
  {\bibfnamefont{C.}~\bibnamefont{Gies}},\ }%
  \bibfield{journal}{%
  \Doi{10.1021/nl500595u}{\bibinfo {journal} {Nano Letters}}\ }%
  \textbf{\bibinfo {volume} {14}},\ \bibinfo {pages} {3743} (\bibinfo {year}
  {2014}),\
  \Eprint{http://arxiv.org/abs/http://pubs.acs.org/doi/pdf/10.1021/nl500595u}{%
http://pubs.acs.org/doi/pdf/10.1021/nl500595u},\
  \url{http://pubs.acs.org/doi/abs/10.1021/nl500595u}%
  \bibAnnoteFile{NoStop}{acs_nanolett}%
\end{thebibliography}%

\end{document}


\title{Supplementary Information for:\\Phase Diagram of Electron Doped Dichalcogenides}

\author{M. R{\"o}sner}
\email{mroesner@itp.uni-bremen.de}
\affiliation{Institut f{\"u}r Theoretische Physik, Universit{\"a}t Bremen, Otto-Hahn-Allee 1, 28359 Bremen, Germany}
\affiliation{Bremen Center for Computational Materials Science, Universit{\"a}t Bremen, Am Fallturm 1a, 28359 Bremen, Germany}

\author{S. Haas}
\affiliation{Department of Physics and Astronomy, University of Southern California, Los Angeles, CA 90089-0484, USA}

\author{T. O. Wehling}
\affiliation{Institut f{\"u}r Theoretische Physik, Universit{\"a}t Bremen, Otto-Hahn-Allee 1, 28359 Bremen, Germany}
\affiliation{Bremen Center for Computational Materials Science, Universit{\"a}t Bremen, Am Fallturm 1a, 28359 Bremen, Germany}

\date{\today}

\maketitle

\begin{figure}[h]

  \psfrag{wph (cm-1)}[c][][0.9]{\textsf{$\omega_\text{ph}$ (cm$^{-1}$)}}
  \psfrag{electron doping x}[c][][0.8]{\textsf{electron doping $x$}}

  \psfrag{Phonon DOS (a.u.)}[][][0.9]{\textsf{Phonon DOS (a.u.)}}
  \psfrag{a2 F(w) (a.u.)}[][][0.9]{ \textsf{$\alpha^2 F(\omega_\text{ph})$ (a.u.)}}
  \psfrag{L(x)}[][][0.9]{$\lambda(x)$}
  \psfrag{wlog(x)}[][][0.9]{\textcolor{myRed}{$\omega_{log}(x)$}}

  \psfrag{x = 0.125}[c][][0.8]{$\quad x=0.125$}
  \psfrag{x = 0.112}[c][][0.8]{$\quad x=0.112$}
  \psfrag{x = 0.100}[c][][0.8]{$\quad x=0.100$}
  \psfrag{x = 0.087}[c][][0.8]{$\quad x=0.087$}
  \psfrag{x = 0.075}[c][][0.8]{$\quad x=0.075$}
  
  \epsfig{file=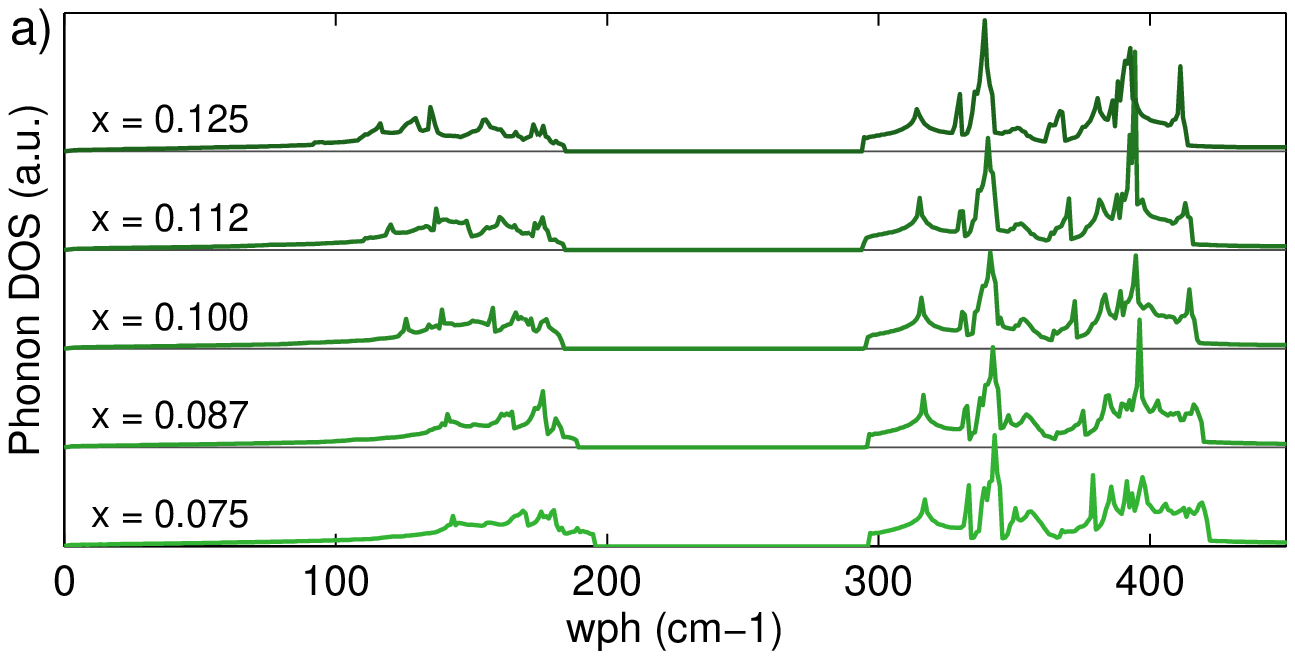,width=7.5cm}
  
  \caption{(Color online) Phonon density of states of $\rm MoS_2$ in the SC phase for different doping levels.}
  \label{dynamics}
  
\end{figure}

Fig. \ref{dynamics} shows the phonon density of states of $\rm MoS_2$ for different doping levels in the regime of the SC phase. While the  high-energy optical branches ($\sim 300-500\,\text{cm}^{-1}$) lead to the strongest peaks in the phonon density of states, it turns out that they do not contribute significantly to the formation of the SC condensate. As it can be seen in the Eliashberg function shown in Fig. 2 of the main text.

\begin{figure}[h]

  \psfrag{Tc (K)}[][][0.9]{\textsf{critical temperature, $T_c(x)$ (K)}}
  \psfrag{a}[][][0.9]{\textcolor{myRed}{\textsf{lattice distortion, $\alpha(x)$} ($^\circ$)}}

  \psfrag{electron doping n2D (1014 cm-2)}[c][][0.9]{\textsf{electron doping $n_{\text{2D}}$ ($10^{14}$ cm$^{-2}$)}}
  \psfrag{electron doping x}[c][][0.9]{\textsf{electron doping $x$}}
  
  \psfrag{semi-metal}[][][0.8]{\textsf{metal}}
  \psfrag{superconductor}[][][0.8]{\textsf{superconductor}}
  \psfrag{CDW}[][][0.8]{\textsf{CDW}}
  
  \psfrag{deformation}[l][l][0.7]{\textsf{$\alpha(x)$}}
  \psfrag{mu=0.05}[l][l][0.7]{\textsf{$\mu^* = 0.05$}}
  \psfrag{mu=0.15}[l][l][0.7]{\textsf{$\mu^* = 0.15$}}
  \psfrag{mu=0.25}[l][l][0.7]{\textsf{$\mu^* = 0.25$}}
  \psfrag{CDW formation}[l][l][0.7]{\textsf{CDW formation}}
  \psfrag{energy}[l][l][0.7]{\textsf{energy}}
  \psfrag{exp. data}[l][l][0.7]{\textsf{exp. data}}
  
  \psfrag{a05877}[l][l][0.7]{\textsf{$a=3.110$\AA}}
  \psfrag{a05900}[l][l][0.7]{\textsf{$a=3.122$\AA}}
  \psfrag{a05922}[l][l][0.7]{\textsf{$a=3.134$\AA}}
  \psfrag{a05944}[l][l][0.7]{\textsf{$a=3.146$\AA}}
  
  \epsfig{file=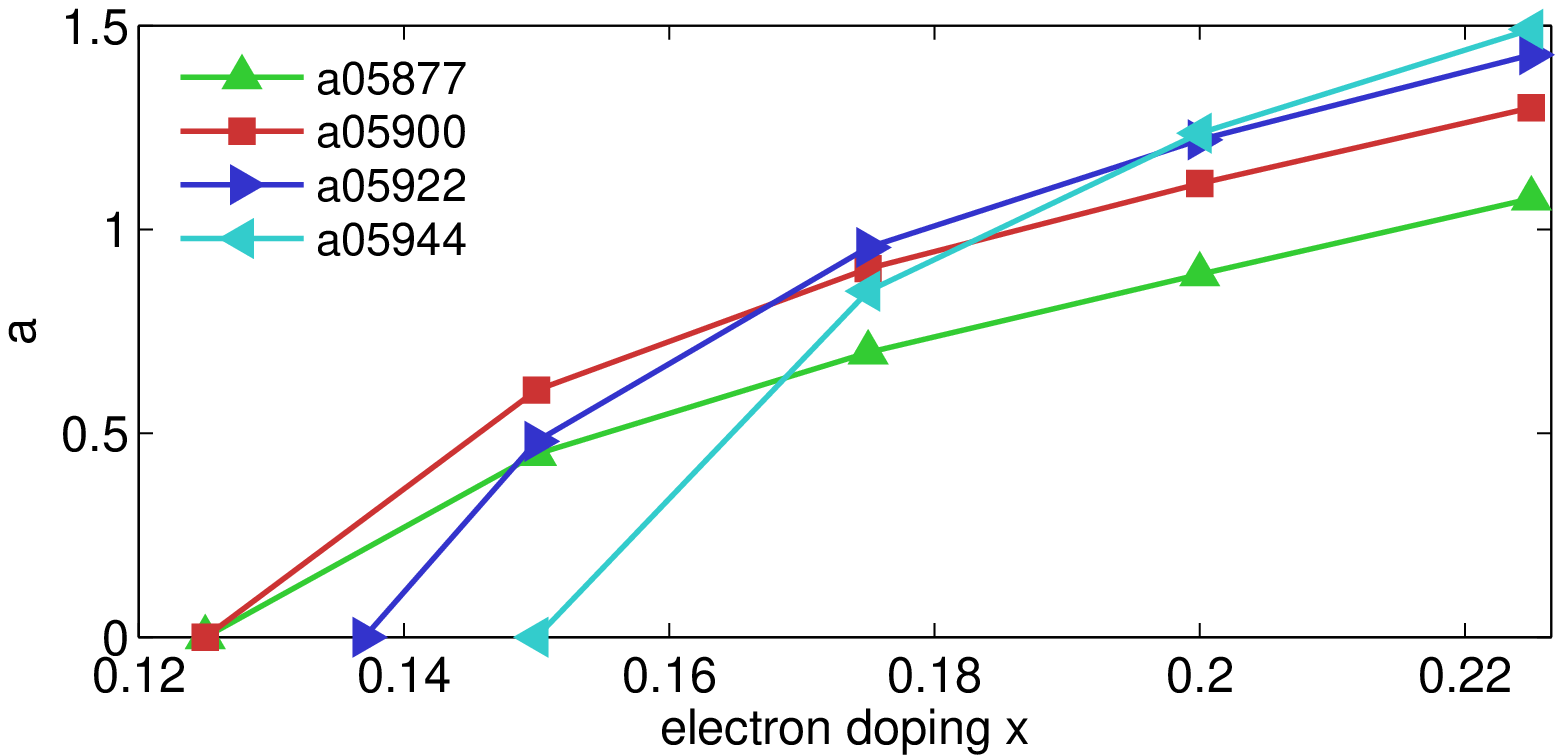,width=7.5cm}
  
  \caption{(Color online) Lattice distortion angle $\alpha$ upon CDW formation in dependence of the electron doping for different lattice constants.}
  \label{dopingdependence}

\end{figure}

Fig. \ref{dopingdependence} shows the lattice distortion upon CDW formation in dependence of the electron doping for different lattice constants. The red curve ($a = 3.122\,$\AA) is the same as in Fig. 3 in the main text. Here we provide the corresponding data for a smaller ($a = 3.110\,$\AA) as well as for two bigger ($a = 3.134\,$\AA and $a = 3.146\,$\AA) lattice constants. The differences between these values are less than $1\%$. However, as it can be seen in Fig. \ref{dopingdependence} the critical concentration for the inset of the CDW changes from $x_c \approx 0.14$ to $x_c \approx 0.17$ upon increasing the lattice constant. This is a change of more than $17\%$. The high doping behavior changes as well and tends to bigger distortion angles with an increasing lattice constant. Indeed, the dependence of the $x_c$ to the lattice constant is bigger, than the high doping behavior, since there seems to be a saturating nature of the lattice distortion which will be reached for every lattice constant.